\documentclass[twocolumn,prl,showpacs,multicol,amsmath,amssymb]{revtex4-1}
\usepackage{graphicx}
\usepackage{epstopdf}
\newcommand{\be}{\begin{equation}}
\newcommand{\ee}{\end{equation}}

\newcommand{\bea}{\begin{eqnarray}}
\newcommand{\eea}{\end{eqnarray}}
\newcommand{\bd}{\begin{displaymath}}
\newcommand{\ed}{\end{displaymath}}
\newcommand{\ba}{\begin{array}}
\newcommand{\ea}{\end{array}}
\newcommand{\bi}{\begin{itemize}}
\newcommand{\ei}{\end{itemize}}
\newcommand{\bc}{\begin{center}}
\newcommand{\ec}{\end{center}}
\newcommand{\bfl}{\begin{flushleft}}
\newcommand{\efl}{\end{flushleft}}
\newcommand{\bfr}{\begin{flushright}}
\newcommand{\efr}{\end{flushright}}

\def\Ce{\rm{CeB$_6$}}
\def\U{\rm{URu$_2$Si$_2$}}

\def\ket#1{\left\vert #1 \right\rangle}

\def\bk{{\bf k}} \def\bq{{\bf q}} 
\def\bQ{{\bf Q}}

\def\6{\partial}

\def\bra{\langle}
\def\ket{\rangle}
\def\={\!\!\!&=&\!\!\!}
\def\+{\!\!\!&&\!\!\!+~}
\def\-{\!\!\!&&\!\!\!-~}

\begin{document}
\date{\today}
\title{Spin Exciton Formation inside  the Hidden Order Phase  of \Ce}

\author{Alireza Akbari}
\author {Peter Thalmeier}

\affiliation{Max Planck Institute for the  Chemical Physics of Solids, D-01187 Dresden, Germany}

\begin{abstract}
The heavy fermion metal \Ce~ exhibits hidden order of antiferroquadrupolar  (AFQ) type below $T_Q=3.2\mbox{K}$ 
and subsequent antiferromagnetic (AFM) order at  $T_N=2.3\mbox{K}$. 
It was interpreted as ordering of the quadrupole and dipole moments of a $\Gamma_8$ quartet of localised Ce $4f^1$ electrons. This  established picture has been profoundly shaken by recent inelastic neutron scattering\cite{friemel:11} that found the evolution of a feedback spin exciton resonance within the hidden order phase at the AFQ wave vector which is stabilized by the AFM order. We develop an alternative theory based on a fourfold degenerate Anderson lattice model, including both order parameters as particle-hole condensates of itinerant heavy quasiparticles. This explains in a natural way the appearance of the spin exciton resonance and the momentum dependence of its spectral weight, in particular around the AFQ vector and its rapid disappearance in the disordered phase. Analogies to the feedback effect in   unconventional heavy fermion superconductors are pointed out.
\end{abstract}

\pacs{71.27.+a, 75.30.Mb, 75.40.Gb}

\maketitle

In strongly correlated f-electron metals the investigation of hidden order (HO) of unconventional non-magnetic type is a topic of central importance\cite{kuramoto:09}. The most prominent and most investigated heavy fermion compounds that exhibit HO at low temperatures are \U~ and \Ce~ which have tetragonal (D$_{4h}$) or cubic (O$_h$) structure respectively.  Two issues arise in the context of hidden order: Firstly, which symmetry is broken in the HO phase and to which irreducible representation the order parameter belongs. Secondly, should the ordering be described as appearance of spontaneous long range correlation between local f-electron degrees of freedoms, i.e. f-electron multipoles, or should HO rather be described as condensation of itinerant heavy particle-hole pairs with a nontrivial orbital structure. These opposite perspectives have prevented a clear identification of the HO in \U~ until present. 

On the other hand since the work of Ohkawa\cite{ohkawa:85} the HO in \Ce~ which appears at T$_Q$=3.2 K has always been taken granted as a paradigm of the localised HO picture. In subsequent work along this line\cite{shiina:97,shiina:98} it was clarified that the primary HO parameter is of the two-sublattice antiferroquadrupolar (AFQ) $\Gamma^+_5$ type $(O_{yz},O_{zx},O_{xy})$   with wave vector $\bQ'=(\frac{1}{2},\frac{1}{2},\frac{1}{2})$  in r.l.u.(R-point) which is nearly degenerate with an antiferrooctupolar (AFO) $\Gamma_2^-$ $(T_{xyz})$ order parameter which is strongly induced in an external field. Here $\pm$ denotes the parity with respect to time reversal. The hidden multipolar order parameters are supported by the fourfold degenerate 4f crystalline electric field (CEF) ground state $\Gamma_8$. This localized scenario explains a large body of experimental results, including the field dependent increase and anisotropy of the critical temperature and the field induced Bragg peaks\cite{erkelens:87}  at $\bQ'$ and NMR line shifts\cite{shiina:98}
although there is no macroscopic symmetry breaking observed \cite{amara:12}. A further important support for this picture comes from the predicted rapid field induced increase of the secondary octupole order parameter\cite{shiina:97} which was directly confirmed by RXS experiments\cite{matsumura:09}. At temperatures below T$_N$=2.3 K finally \Ce~ develops antiferromagnetism (AFM) with $\bQ=(\frac{1}{4},\frac{1}{4},0)$ ($\Sigma$ or S -point) that coexists with AFQ order.
Important information on HO may also be gained from the magnetic excitation spectrum. For finite fields that stabilizes the AFQ/AFO HO it was investigated within generalized Holstein-Primakoff and  random phase approximation (RPA) approaches\cite{shiina:03,thalmeier:03}. Both lead to multipolar excitation bands in the range $1-2.5$ meV and for finite applied field\cite{bouvet:93} their salient features agree with experimental results from inelastic neutron scattering (INS).
%
\begin{figure*}[t]
\centering
\includegraphics[width=0.28\linewidth]{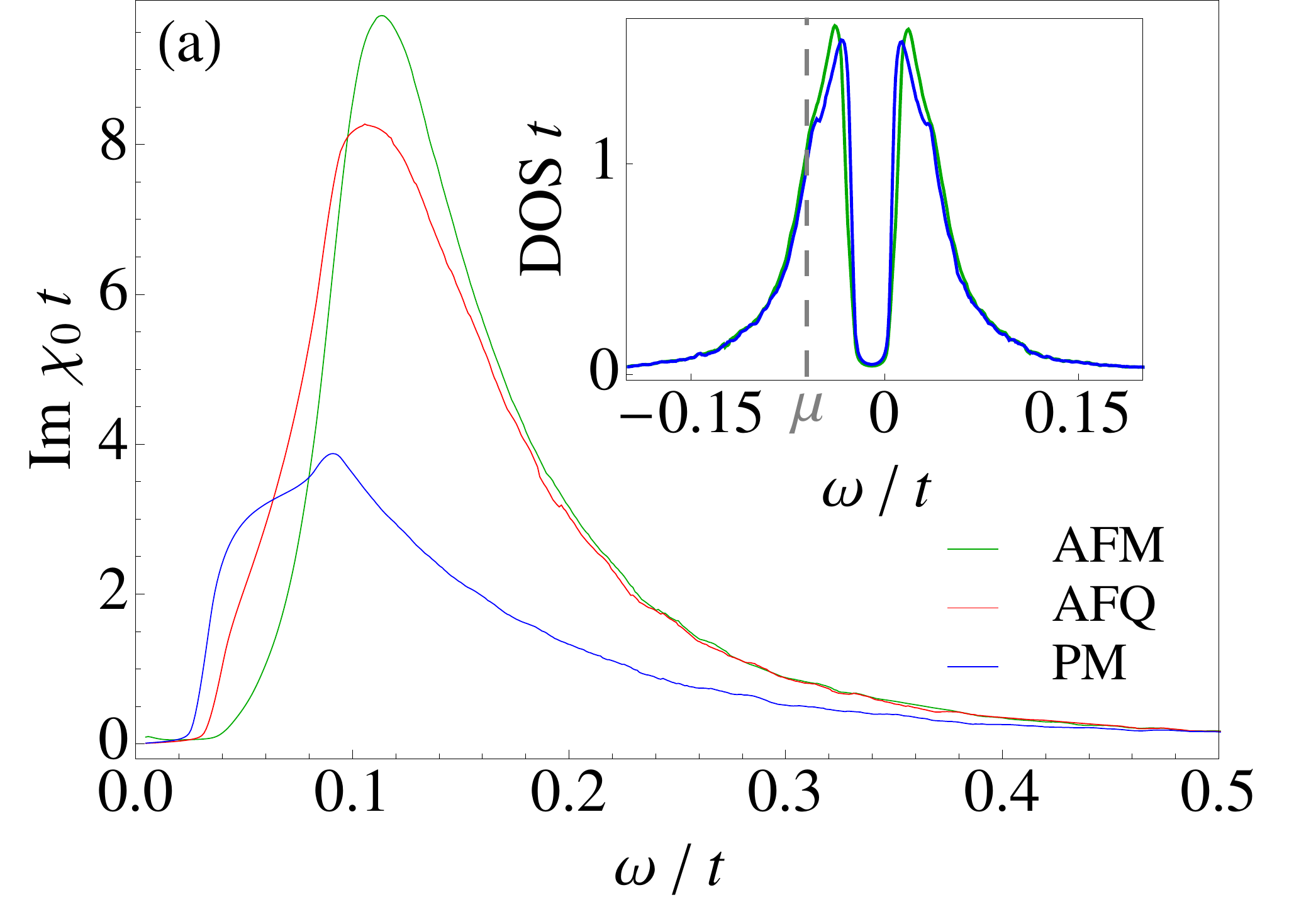}
\includegraphics[width= 0.28\linewidth]{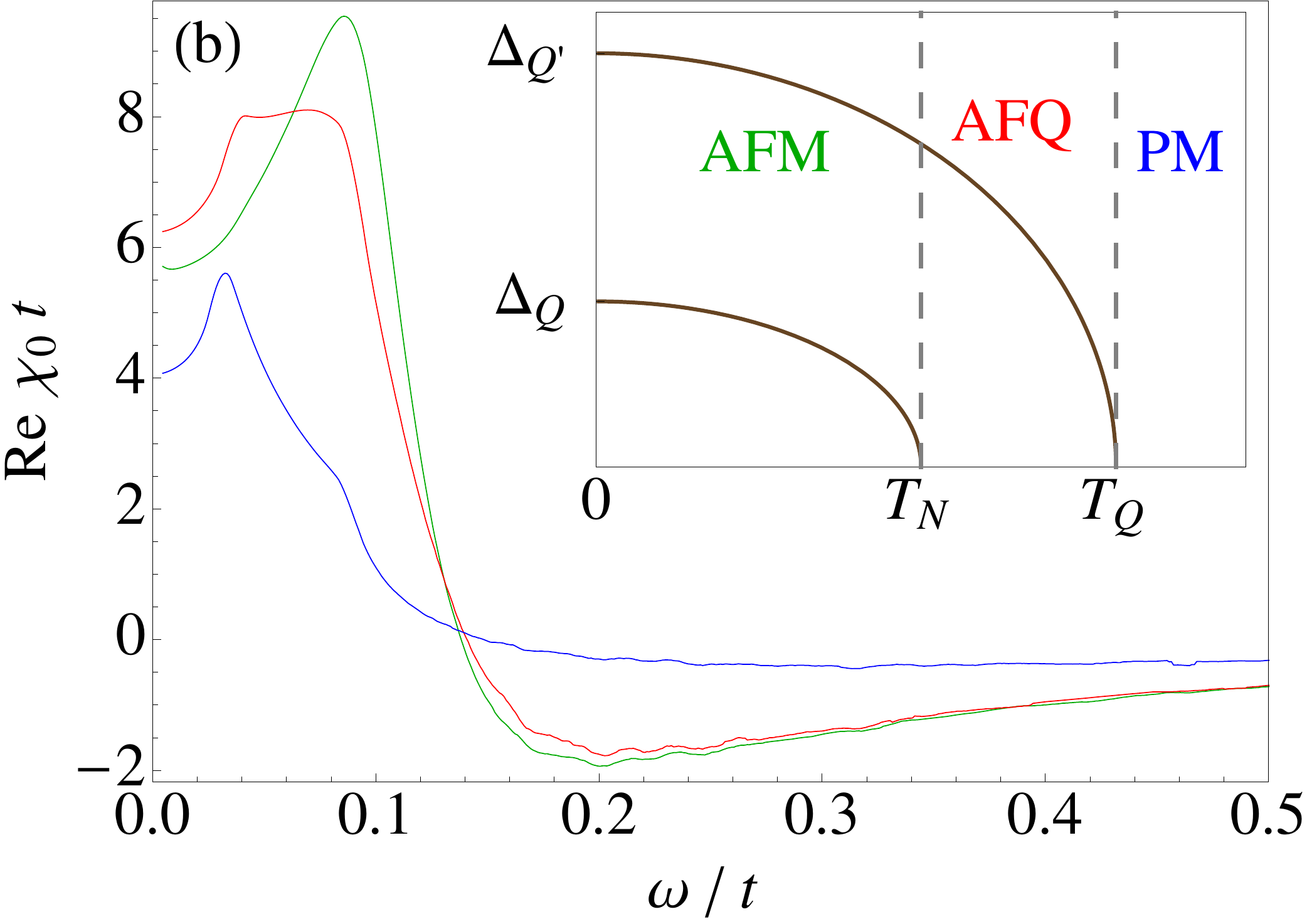}
\includegraphics[width= 0.28\linewidth]{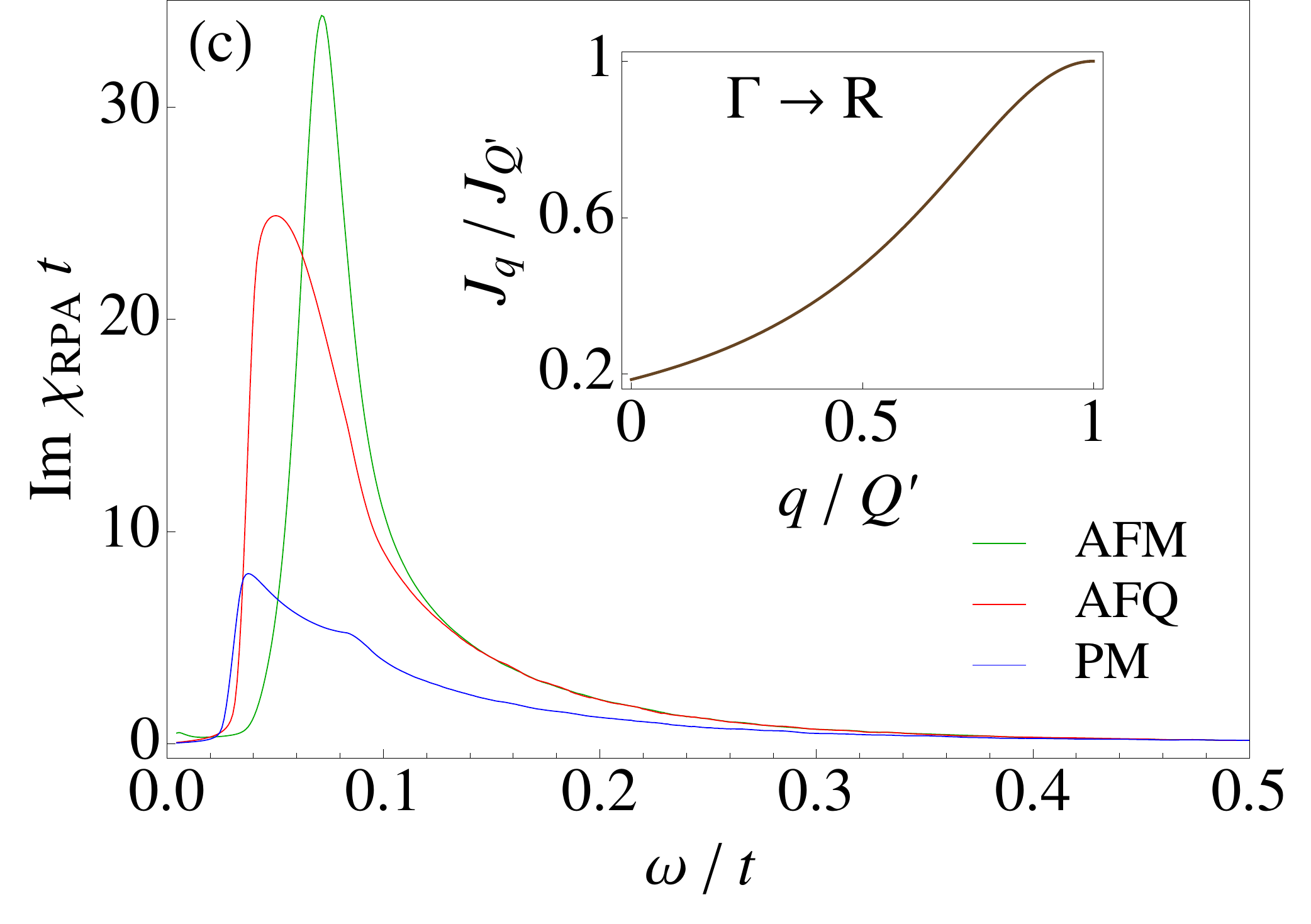}
\caption{
(Color online) Non-interacting susceptibility at the R-point $(\frac{1}{2},\frac{1}{2},\frac{1}{2})$: (a) imaginary part and 
(b) real part. Inset of (a) shows the quasiparticle DOS in PM and coexisting AFQ/AFM (T=$6\times 10^{-3}t$) phase where $\mu =-0.06t$ is the chemical potential. Inset of (b) gives the schematic  temperature dependence of order parameters.
quasiparticle model parameters: $t=22.4$ meV; $\tilde{V}=0.3t$; $\tilde{\epsilon}_f=-0.01t$; gap parameters: $\Delta_{\bQ'}=0.015t$; $\Delta_{\bQ'}=0.005t$
(c) Imaginary part of RPA susceptibility at R-point 
(inset shows the model for the quasiparticle interaction $J_\bq$ along $\Gamma$R direction with $J_{\bQ'}=0.1t$).}
\label{fig:Fig1}
\end{figure*}
%
In the numerous theoretical investigations of HO in \Ce~the localised 4f approach was chosen and itinerant 4f character was completely neglected. This seems surprising because \Ce~is a prominent example of a heavy fermion metal with one of the heaviest masses reported ($m^*/m \geq 17 $)\cite{joss:87} and the Ce- dilute La- substitutes\cite{hewson:93} being the standard case of Kondo resonance dominated local Fermi liquids with all the typical Kondo anomalies identified there. In fact the estimated Kondo temperature of the concentrated \Ce~ from quasielastic neutron scattering\cite{loewenhaupt:85} is $T^*\simeq$ 4.5 K  which is of the same order as $T_Q$ and T$_N$. Therefore one question is whether the HO physics of \Ce~ can be completely explained within the conventional localized 4f approach.

Recent zero field high resolution INS experiments  by Friemel et al\cite{friemel:11} have indeed seriously questioned the standard picture and found intriguing new evidence that the dynamical magnetic response in the HO phase cannot be understood in the localized approach and, as in \U,~requires taking into account the itinerant quasiparticle nature of f electrons.  It was found that the low temperature  magnetic response within the HO phase is determined by a pronounced {\it feedback effect},  i.e. a modification of magnetic spectral properties due to the appearance of order parameters:
 i) Below $T_N$ a spin gap opens for low energies and spectral weight from the quasielastic region \cite{mishchenko:98} is shifted to higher energies forming a pronounced resonance at $\bQ'$ with peak position $\omega_r\simeq 0.5$ meV. ii) Using the single-particle charge gap $2\Delta \simeq 1.2$ meV in the HO phase from point-charge spectroscopy\cite{kunii:87} $\omega_r/2\Delta = 0.42 < 1$ is fulfilled showing that the resonance is indeed split off from the continuum. iii) The resonance appears mainly at the AFQ $\bQ' $ but not at the AFM \bQ~ vector and shows no dispersion. Its intensity decreases rapidly when approaching T$_N$ from below in an order-parameter like fashion. These characteristics of the magnetic spectrum in \Ce~ do not suggest a spin wave origin but rather are reminiscent of spin exciton resonances observed before in Fe-pnictide\cite{lumsden:09} and heavy fermion superconductors\cite{stock:08,stockert:11} as well as Kondo insulators\cite{nemkovski:07,akbari:09}. The results of Ref.\onlinecite{friemel:11} are the first clearcut example of the feedback spin exciton appearing within the AFQ HO phase. This proves that the localized 4f-scenario for \Ce~ is not adequate to explain its intriguing low energy spin dynamics and its momentum dependence.

In this Letter we therefore propose and explore an alternative route of theoretical modeling. We start from the central idea that the AFQ and AFM order parameters are to be described as particle hole condensates in the itinerant heavy quasiparticle picture. The latter is obtained from a microscopic fourfold ($\Gamma_8$-type) degenerate Anderson lattice model. It  includes both twofold pseudo-spin ($\sigma =\uparrow,\downarrow$) and twofold pseudo-orbital ($\tau=\pm$) degeneracies of the hybridizing conduction (c) and 4f electron (f) in the $\Gamma_8$ CEF ground state according to
\begin{eqnarray}
{\cal H}
&=&
\sum\limits_{{\bf k},m}
\left[\epsilon^c_{{\bf k}}c_{{\bf k}m
}^{\dagger}c_{{\bf k}m }
+
\epsilon^f_{{\bf k}} f_{{\bf k}m} ^{\dagger}f_{{\bf k}m}
+V_{{\bf k}}\left( c_{{\bf k},m }^{\dagger}f_{{\bf k}m}
+h.c.\right)\right]
 \nonumber \\
&&+\sum\limits_{i,m,n} U_{ff}f_{ i m }^{\dagger}f_{i n }f_{i n}^{\dagger}f_{i  m }.
\label{eq:HAL}
\end{eqnarray}
where $m=(\tau,\sigma)$ represents the fourfold $\Gamma_8$ degeneracy. Here $c_{{\bf k}m
}^{\dagger}$ creates an conduction electron in the channel with corresponding $\Gamma_8$ symmetry
and wave vector \bk.
Furthermore, $\epsilon^c_{{\bf k}}$ and $\epsilon^f_{{\bf k}}=\epsilon^f$  are effective tight binding
dispersions of the conduction band and the atomic $f$ level position respectively. For the former 
we restrict to the next neighbor hopping ($t$), i.e., $\epsilon^c_{{\bf k}}=2t\sum_{n}\cos k_n$ $(n=x,y,z)$ which leads
naturally to the AFQ ordering vector $\bQ' $.
Furthermore $f_{{\bf k}m} ^{\dagger}$ creates the f electron with momentum \bk, and $U_{ff}$ is its on-site Coulomb 
repulsion. Finally $V_{{\bf  k}}$ is the  hybridization energy between the lowest 4f doublet
and conduction bands which contains in principle the effect of spin orbit and CEF but is taken as constant $V_{{\bf  k}}=V$ 
here.

In  the limit $U_{ff}\rightarrow \infty$ double occupation  of the f-states are excluded,
this is achieved by using the auxiliary boson $b_i$ at each site $i$,  with the  constraint
$b_i^{\dagger}b_i+\sum_{m}f^{\dagger}_{im}f_{im}=1$. In the mean field (MF) approximation
($r=\langle b_i \rangle=\langle b_i^{\dagger} \rangle$)
diagonalization leads to hybridized quasiparticle bands\cite{hewson:93}.
They are determined by renormalized f level $\tilde{\epsilon}^{f}_{{\bf  k}}=\epsilon^{f}_{{\bf  k}}+\lambda$ 
and effective (reduced) hybridization $\tilde{V} _{\bf  k}=rV _{\bf  k}$.
Minimizing the MF ground state energy leads to selfconsistent
equations for $r,\lambda$.

The AFQ and AFM order parameters with wave vectors \bQ~ and $\bQ' $ respectively contribute extra 
MF terms
\bea
{\cal H}_{AFQ}&=&\sum\limits_{{\bf k}\sigma} 
\Delta_{{\bf Q}^\prime}(f_{{\bf k},+\sigma }^{\dagger}f_{{\bf k}+{\bf Q}^\prime-\sigma }+f_{{\bf k}, -\sigma}^{\dagger}f_{{\bf k}+{\bf Q}^\prime,+\sigma}),
\nonumber\\
{\cal H}_{AFM}&=&\sum\limits_{{\bf k} \tau} \Delta_{\bf Q}
(f_{{\bf k} \tau\uparrow }^{\dagger}f_{{\bf k}+{\bf Q} \tau\downarrow }+f_{{\bf k} \tau\downarrow }^{\dagger}f_{{\bf k}+{\bf Q} \tau\uparrow }).
\label{eq:OP}
\eea
Our emphasis in this work is on the feedback effect, i.e. the effect of  the gap opening within the HO phase on the magnetic response. Therefore we do not attempt  a microscopic calculation to derive these order parameters and their temperature dependence. We include them as symmetry breaking molecular field terms in the Hamiltonian and take a generic empirical temperature dependence. The MF Hamiltonian ${\cal H}_{MF}$ obtained from Eq.~(\ref{eq:HAL}) is diagonalized by the unitary transformation 
\bea
f_{{\bf k}m }=u_{+, {\bf k}} a_{+,{\bf k}m }+u_{-, {\bf k}} a_{-,{\bf k}m }
\nonumber\\
c_{{\bf k}m }=u_{-, {\bf k}} a_{+,{\bf k}m }-u_{+, {\bf k}} a_{-,{\bf k}m }.
\eea
where 
$                              
2u_{\pm, {\bf k}}^2 =
1\pm (\epsilon^{c}_{{\bf  k}}-\tilde{\epsilon}^{f}_{{\bf  k}})/\sqrt{(\epsilon^{c}_{{\bf  k}}
-\tilde{\epsilon}^{f}_{{\bf  k}})^2+4\tilde{V}^2_{{\bf  k}}}
$, leading to 
${\cal H}_{MF}= \sum\limits_{i,{\bf k},m }E^{\alpha}_{{\bf k}}a^{\dagger}_{\alpha,{\bf k}m}a_{\alpha,{\bf k}m}+\lambda(r^2-1)$,
where 
$
E^{\pm} _{\bf  k}=\frac{1}{2}
[
\epsilon^{c}_{{\bf  k}}+\tilde{\epsilon}^{f}_{{\bf  k}}\pm\sqrt{(\epsilon^{c}_{{\bf  k}}
-\tilde{\epsilon}^{f}_{{\bf  k}})^2+4\tilde{V}^2_{{\bf  k}}}
]
$
are the pair ($\alpha =\pm$) of hybridized quasiparticle ($a_{\alpha{\bf k}m}$) bands, each fourfold (m=1-4) degenerate.
Here $\tilde{V}^2 _{\bf  k}=V^2 _{\bf  k}(1-n_f)$ denotes the effective hybridization obtained by
projecting out double occupancies. Due to $1-n_f\ll 1$ $\tilde{V} _{\bf  k}$ is strongly reduced
with respect to the single particle $V _{\bf  k}$ which leads to the large quasiparticle mass.
Introducing new Nambu operators as 
$
\psi^{\dagger}_{\bf k}=(C^{\dagger}_{\bf k},C^{\dagger}_{{\bf k}+{\bf Q}^\prime},C^{\dagger}_{{\bf k}+{\bf Q}})
$
where
$
C^{\dagger}_{\bf k}=(b^{\dagger}_{+,{\bf k}},b^{\dagger}_{-,{\bf k}})
$
and 
$
b^{\dagger}_{\alpha,{\bf k}}=
(a^{\dagger}_{{\alpha,{\bf k}}+\uparrow},a^{\dagger}_{{\alpha,{\bf k}}+\downarrow},
a^{\dagger}_{{\alpha,{\bf k}}-\uparrow},a^{\dagger}_{{\alpha,{\bf k}}-\downarrow})
$,
we can write the total Hamiltonian ${\cal H}_{tot}={\cal H}_{MF}+{\cal H}_{AFQ}+{\cal H}_{AFM}$ as
\bea\nonumber
{\cal H}_{tot}=\sum\limits_{{\bf k}} 
\hat{\psi}_{{\bf k} }^{\dagger}\hat{\beta}_{{\bf k}}
\hat{\psi}_{{\bf k} }
;\;\;
\hat{\beta}_{{\bf k}}=
\left[
 \begin{array}{ccc}
\hat{E}_{\bf  k}  & \hat{\Delta}_{{\bf Q}^\prime}  & \hat{\Delta}_{\bf Q}\\
 \hat{\Delta}_{{\bf Q}^\prime}  & \hat{E}_{{\bf  k}+{\bf Q}^\prime}  &  0 \\
\hat{\Delta}_{\bf Q} &0 & \hat{E}_{{\bf  k}+{\bf Q}} ,
\end{array}
\right],
\eea
here
$
\hat{E}_{\bf  k} =\hat{\cal E}_{\bf  k}  \otimes  \tau_0 \otimes \sigma_0
$,
$
 \hat{\Delta}_{{\bf Q}^\prime}= \Delta _{{\bf Q}^\prime}( \hat{ \rho}_{{\bf  k},{\bf Q}^\prime} \otimes  
 \hat{\tau}_0 \otimes \hat{\sigma}_x)
$
, and
$
 \hat{\Delta}_{\bf Q}= \Delta _{\bf Q}(\hat{ \rho}_{{\bf  k},{\bf Q}} \otimes  \hat{\tau}_x \otimes  \hat{  \sigma}_0)
$,
where 
$\hat{\cal E}_{\bf  k} $ and $\hat{ \rho}_{{\bf  k},{\bf Q}^\prime}$  are $2\times 2$ matrices in $\alpha=\pm$ space with matrix elements
$
\hat{\cal E}_{\bf  k} ^{\alpha\beta} = \delta_{\alpha\beta} E^{\alpha} _{\bf  k}
$, and
$
 \hat{ \rho}_{{\bf  k},{\bf k}^\prime} ^{\alpha\beta} = u_{\alpha{\bf k}} u_{\beta,{\bf k}+{\bf k}^\prime}
$.
$\sigma_l, \tau_l$ are the Pauli matrices acting in pseudo-spin and pseudo-orbital space, respectively.
  
Defining the  Matsubara GreenÕs function (GF) matrix as 
${\hat G}_{ {\bf  k}}(\tau)=-\langle T \hat{\psi}_{{\bf k} }(\tau)
\hat{\psi}_{{\bf k}}^{\dagger}(0)\rangle$, 
and solving the standard equations of motion,
one can find
$
 {\hat G}_{{\bf k}}(\omega_n)=\left( i\omega_n-\hat{\beta}_{{\bf k}} \right)^{-1}
$
 which can be written as
 \bea
 {\hat G}_{{\bf k}}(\omega_n)=
\left[
 \begin{array}{ccc}
 {\hat G}^0_{{\bf k}} & {\hat G}^0_{{\bf k} ,{\bf k}+{\bf Q}^\prime}   &{\hat G}^0_{{\bf k},{\bf k}+{\bf Q}  }\\
{\hat G}^0_{{\bf k}+{\bf Q}^\prime,{\bf k} }  &  {\hat G}^0_{{\bf k}+{\bf Q}^\prime } &  {\hat G}^0_{{\bf k}+{\bf Q}^\prime,{\bf k}+{\bf Q}  } \\
{\hat G}^0_{{\bf k}+{\bf Q}{\bf k} }  &{\hat G}^0_{{\bf k}+{\bf Q},{\bf k}+{\bf Q}^\prime  } & {\hat G}^0_{{\bf k}+{\bf Q} }
\end{array}
\right],
\eea
here $\hat{G}^0_{\bk}$ is a $8\times 8$ Green's function matrix in $(\alpha,m)$ space.
For the magnetic excitation spectrum we need the dipolar  susceptibility matrix given by
 $
 \chi_{{\bf q}}^{ll^\prime}(t)=-
 \theta(t)
 \bra
 T j_{{\bf q}}^{l}(t)j_{-{\bf q}}^{l^\prime}(0)
 \ket,
 $
 where
$
  j_{{\bf q}}^{l}=\sum\limits_{{\bf k}mm^\prime} 
  f_{{\bf k}+{\bf q}m} ^{\dagger}
  {\hat M}^{l}_{mm^\prime} 
  f_{{\bf k}m^\prime}
$ 
are  the physical magnetic dipole operators ($l,l^\prime=x,y,z$).
In cubic symmetry it is sufficient to calculate $ \chi_{{\bf q}}^{zz}(\omega)$, 
corresponding to\cite{thalmeier:03}  $ {\hat M}^{z}=\frac{7}{6} \hat{\tau}_0 \otimes \hat{\sigma}_z$, 
defining $s= (\alpha,{{\bf  k}+{\bf q}}, m_1 )$ and  $s^\prime=(\alpha^\prime,{{\bf  k}}, m_2) $
one finds
 \bea
 \chi_{0}({\bf q},\omega)&&= \chi_{{\bf q}}^{zz}(\omega)
\propto
 \sum\limits_{\alpha \alpha^\prime{\bf k}m_1m_2} 
   ( \hat{ \rho}_{{\bf  k},{\bf q} } ^{\alpha^\prime \alpha})^2
   \int d\omega^\prime
   \nonumber\\&&
   {\hat G}^{0}_{ss}(\nu+\omega^\prime)      {\hat G}^{0 }_{s^\prime s^\prime }(\omega^\prime)
  \mid _{i\nu\rightarrow \omega+i 0^+}
 \eea
Here the  $\hat{ \rho}_{{\bf  k},{\bf q} } ^{\alpha^\prime \alpha}$ are the  matrix elements of reconstructed quasiparticle states 
in the AFQ/AFM state. They play a similar role as the 'coherence factors' in the spin exciton formation in unconventional superconductors.
The dynamic magnetic susceptibility in RPA has the 
form
\bea
 \chi_{RPA}({\bf q},\omega)=
 [1- J_{{\bf q}}  \chi_{{\bf q}}^{zz}(\omega)]^{-1} \chi_{{\bf q}}^{zz}(\omega),
 \label{eq:RPA}
\eea
where $J_{{\bf q}}$ is the heavy quasiparticle interaction taken diagonal in $(\alpha,m)$ band indices. In principle it is determined by processes beyond the slave boson
MF approximation\cite{riseborough:92}. However as in other spin-exciton theories\cite{akbari:09,eremin:08} we adopt
here an empirical form of Lorentzian type that is peaked at the AFQ ordering vector where the resonance appears.

%
\begin{figure}[t]
\centerline{
\includegraphics[width=0.4\linewidth]{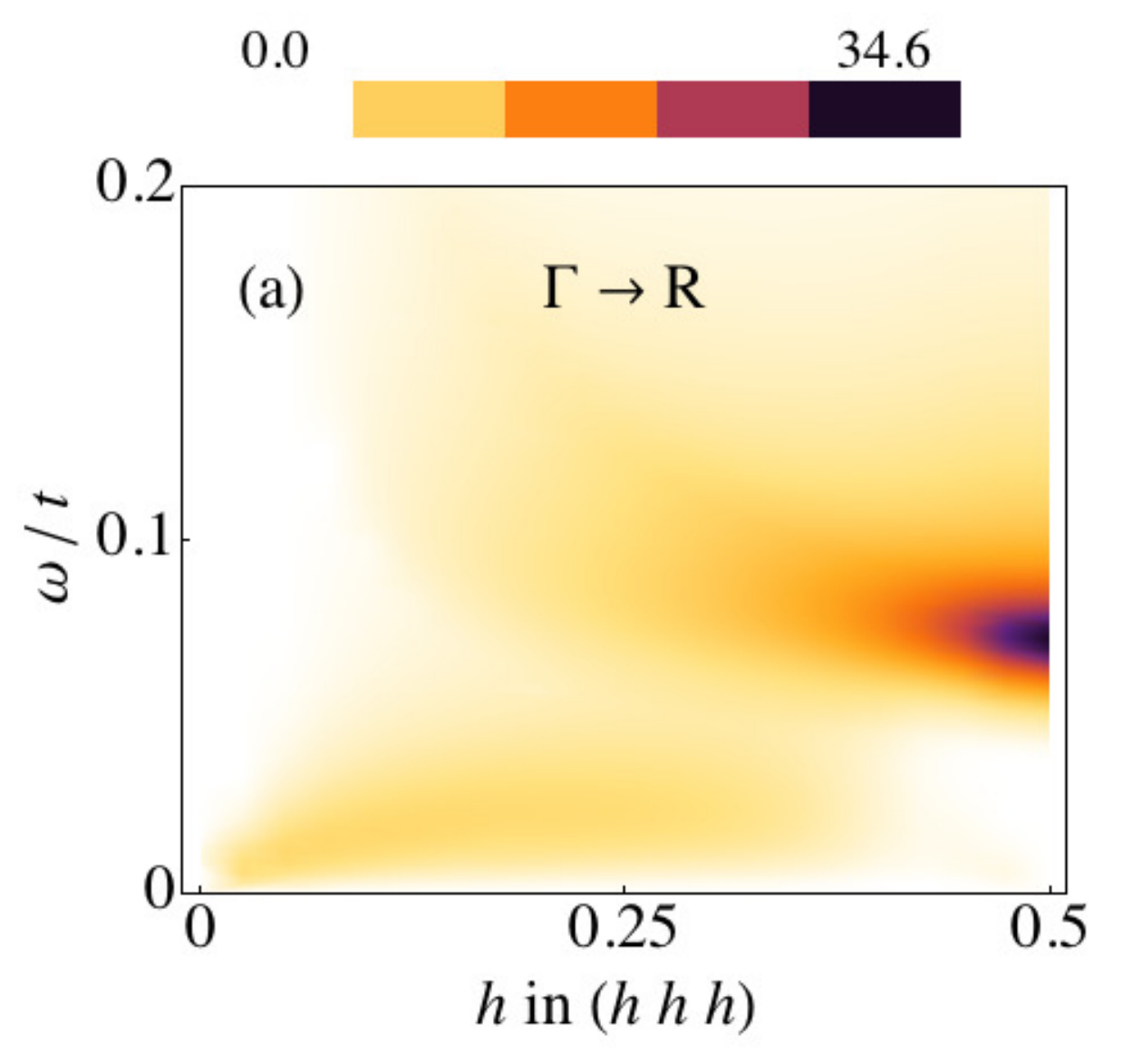}
\includegraphics[width=0.4\linewidth]{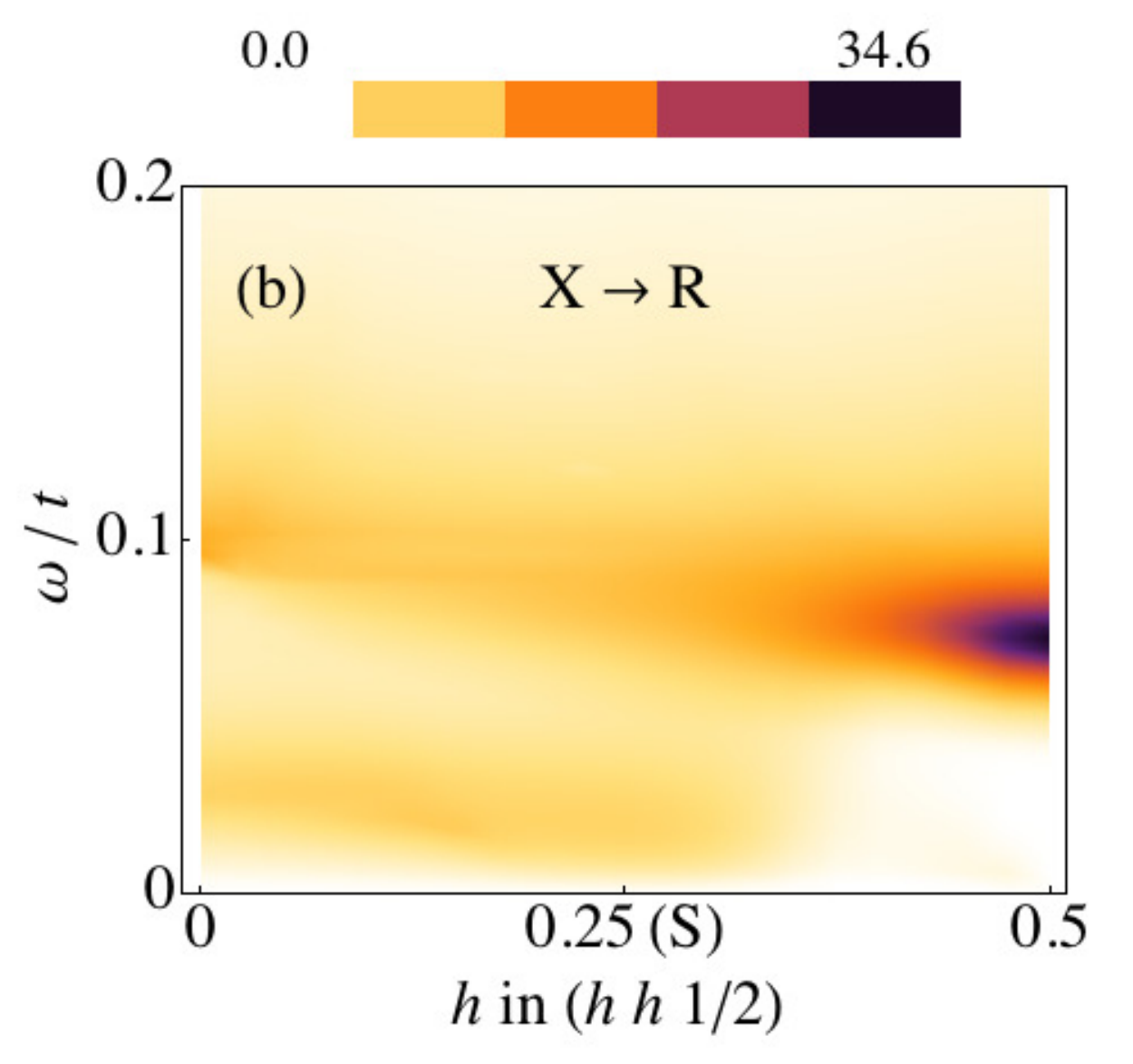}
}
\centerline{
\includegraphics[width=0.4\linewidth]{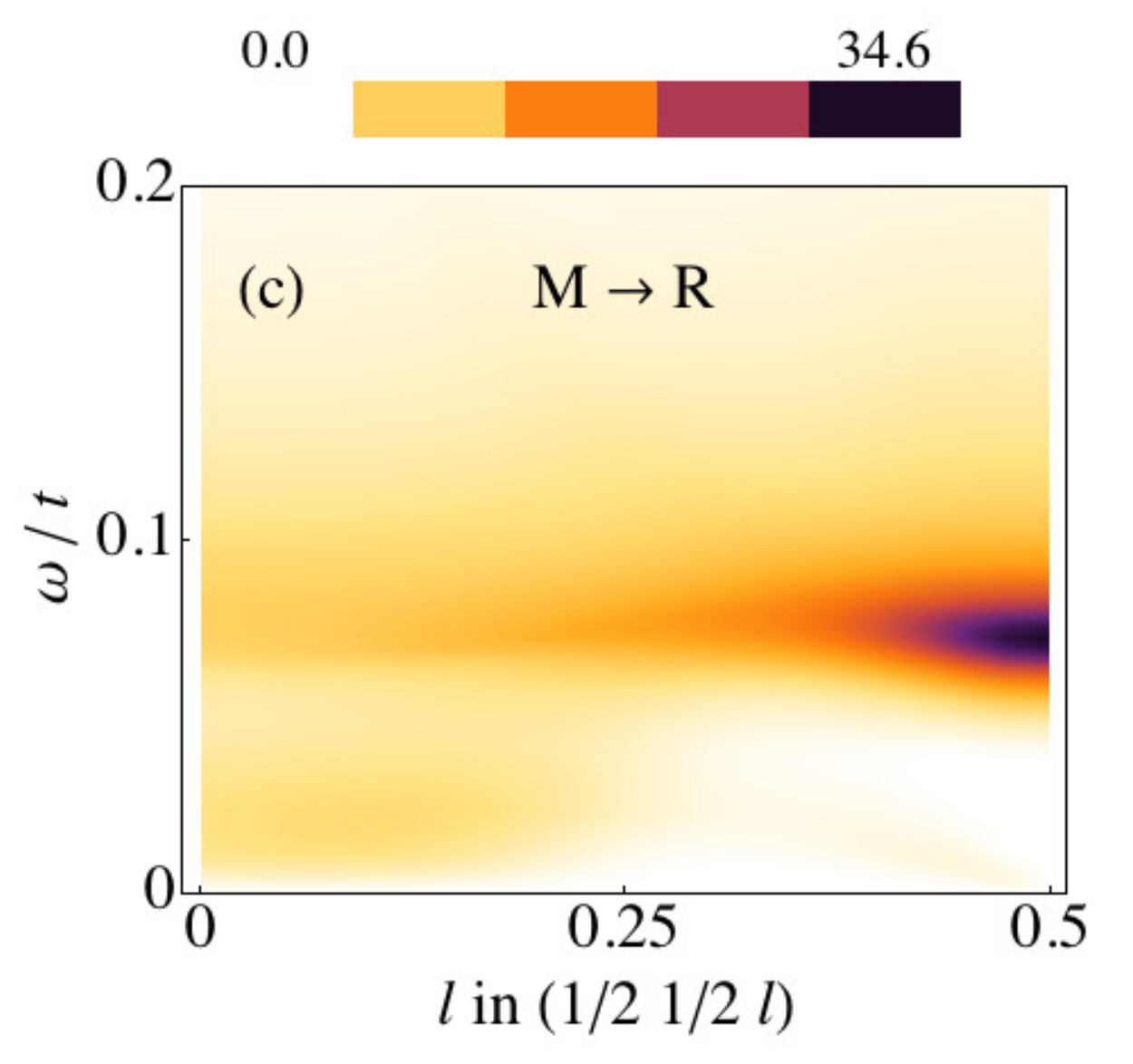}
\includegraphics[width=0.4\linewidth]{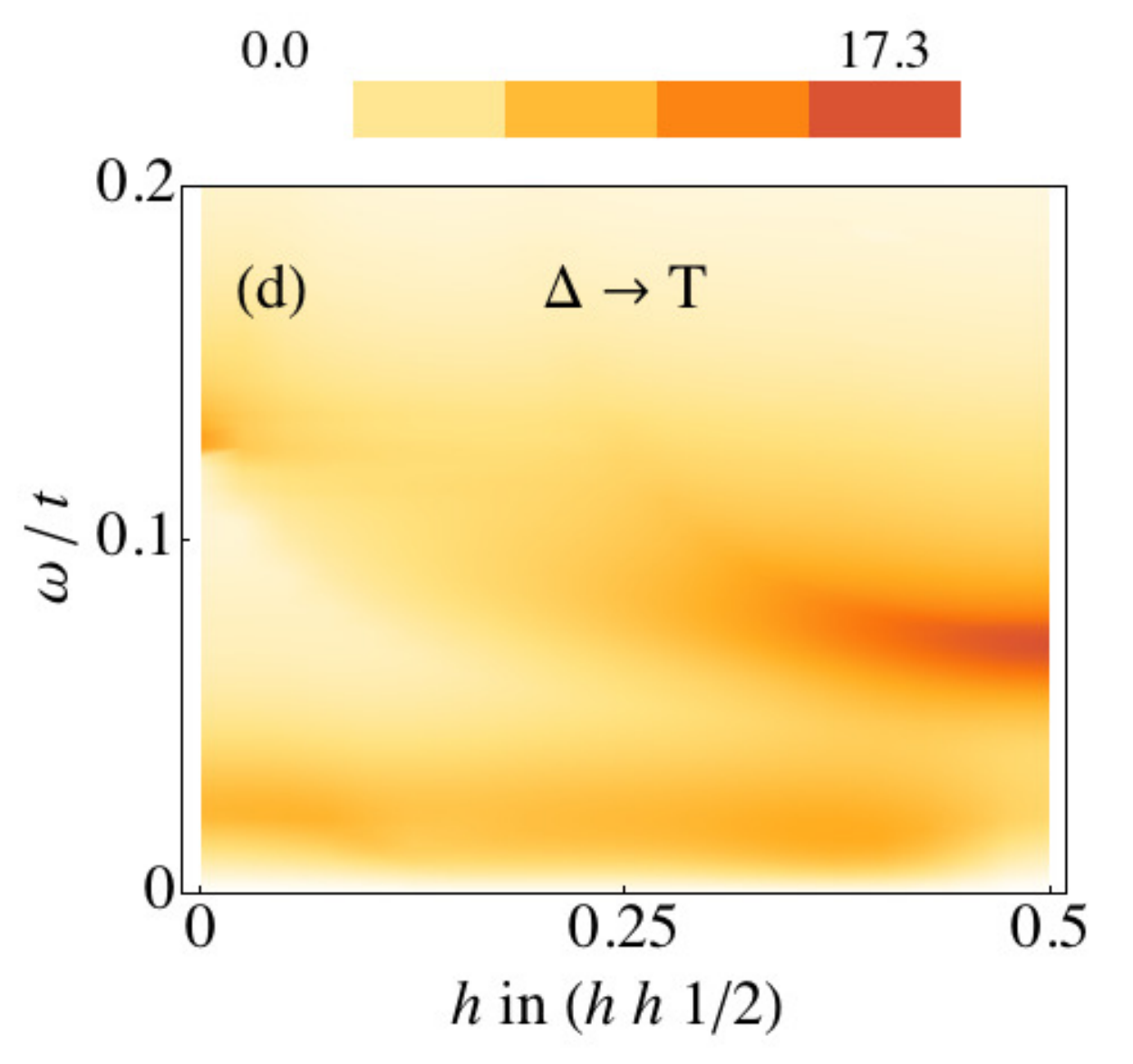}
}
\caption{
(Color online) Contour plot of imaginary part of RPA dynamical
susceptibility (a) from $\Gamma(0~0~0)$ to R$(\frac{1}{2}~\frac{1}{2}~\frac{1}{2})$;
(b)  from $X(0~0~\frac{1}{2})$ to R-point; 
(c) from $X$-point to R-point;
(d)  from $\Delta(0~0~\frac{1}{2})$ to T$(\frac{1}{2}~\frac{1}{2}~\frac{1}{4})$
(note different scale). Resonance is located around R and shows little dispersion.
}
\label{fig:Fig2}
\end{figure}
%
We will now discuss the characteristics of the magnetic excitation spectrum obtained from $ \chi^"_{RPA}({\bf q},\omega)$ and 
show that it explains all the essential experimental features observed in \Ce. In accordance with the heavy quasiparticle mass in this compound the chemical potential is chosen close to the top of the lower quasiparticle band (Fig.~\ref{fig:Fig1}a inset: $\mu= - 0.06t$) where dispersion is flat, leading to a realistic mass enhancement $m^*/m \simeq 20$. All other model parameters are defined in Fig.~\ref{fig:Fig1}.

First the spectrum $\chi^"_{0}({\bf q},\omega)$  of {\it non-interacting} quasiparticles is shown in Fig.\ref{fig:Fig1}a with constant-\bq~ scans for the paramagnetic (PM), AFQ  and coexistent AFQ/AFM phases, respectively. In the PM state the
spectrum exhibits the cf- hybridization gap at the R-point. When the AFQ, AFM order appears their corresponding gaps $\Delta_{\bQ'}$ and  $\Delta_{\bQ}$ push the magnetic response to higher energies. The associated real part in  Fig.\ref{fig:Fig1}b then shows a much enhanced response at these energies. As a consequence the magnetic spectrum for the {\it interacting} quasiparticles may develop a resonance when the real part of the denominator in Eq.~(\ref{eq:RPA})  is driven to zero equivalent to a pole in  $\chi_{RPA}({\bQ'},\omega)$.  Due to the 3D electronic structure  $\chi'_{0}({\bf Q'},\omega)$ will not be singular and the resonance will only appear for $J({\bf Q'})$ larger than a threshold value. The imaginary part is generally non-zero but small leading to a large resonant response at the pole position. The resonance appears in the HO phase when $J_{\bQ^\prime}/t$ lies in a reasonable range such that the pole exists only when the real part is enhanced by the gap formation. Then the resonance condition $1=J_{\bQ^\prime}\chi_{0}({\bQ^\prime},\omega_r)$ is fulfilled only in the AFQ ordered regime. The magnetic spectrum of interacting quasiparticles is shown in Fig.\ref{fig:Fig1}c. It shows indeed a peak appearing in the AFQ phase and a sharp resonant peak at $\omega_r/2\Delta_c = 0.64 $ at low temperature when both gaps are present. Here $\Delta_c = 0.056t$ is the charge gap given in  the inset of Fig.\ref{fig:Fig1}a. This explains the central observation of the R-point resonance in \Ce.

The momentum dependence of the spectrum in the AFQ/AFM phase and in particular the resonance peak is shown in Fig.~\ref{fig:Fig2} as contour plot in the $\bq,\omega$ plane with the wave vector \bq~ chosen along various symmetry directions. 
There are two main characteristics: i) The single-particle spin gap due to the hybridization and enhanced by the AFQ/AFM gap formation appears most prominently  close to the R point and less at other symmetry points like, e.g., T($\frac{1}{2},\frac{1}{2},\frac{1}{4}$). ii) The many-body resonance peak is also strongly constrained to the narrow region around the R-point, partly due to the suppression of the  $\chi'_{0}({\bf q},\omega)$ peak (Fig.~\ref{fig:Fig1}b) when \bq~ moves away from $R(\frac{1}{2},\frac{1}{2},\frac{1}{2})$ and partly due to the 
decrease of $J_\bq$. Both mean that the above condition for the resonance can only be fulfilled in a narrow region around the R-point where it is almost dispersionless. This corresponds exactly to the experimental observation in \Ce~   and similar observations have been made in the Ce-based superconductors \cite{stock:08,stockert:11}.
%
\begin{figure}[t]
\centerline{
\includegraphics[width=0.4\linewidth]{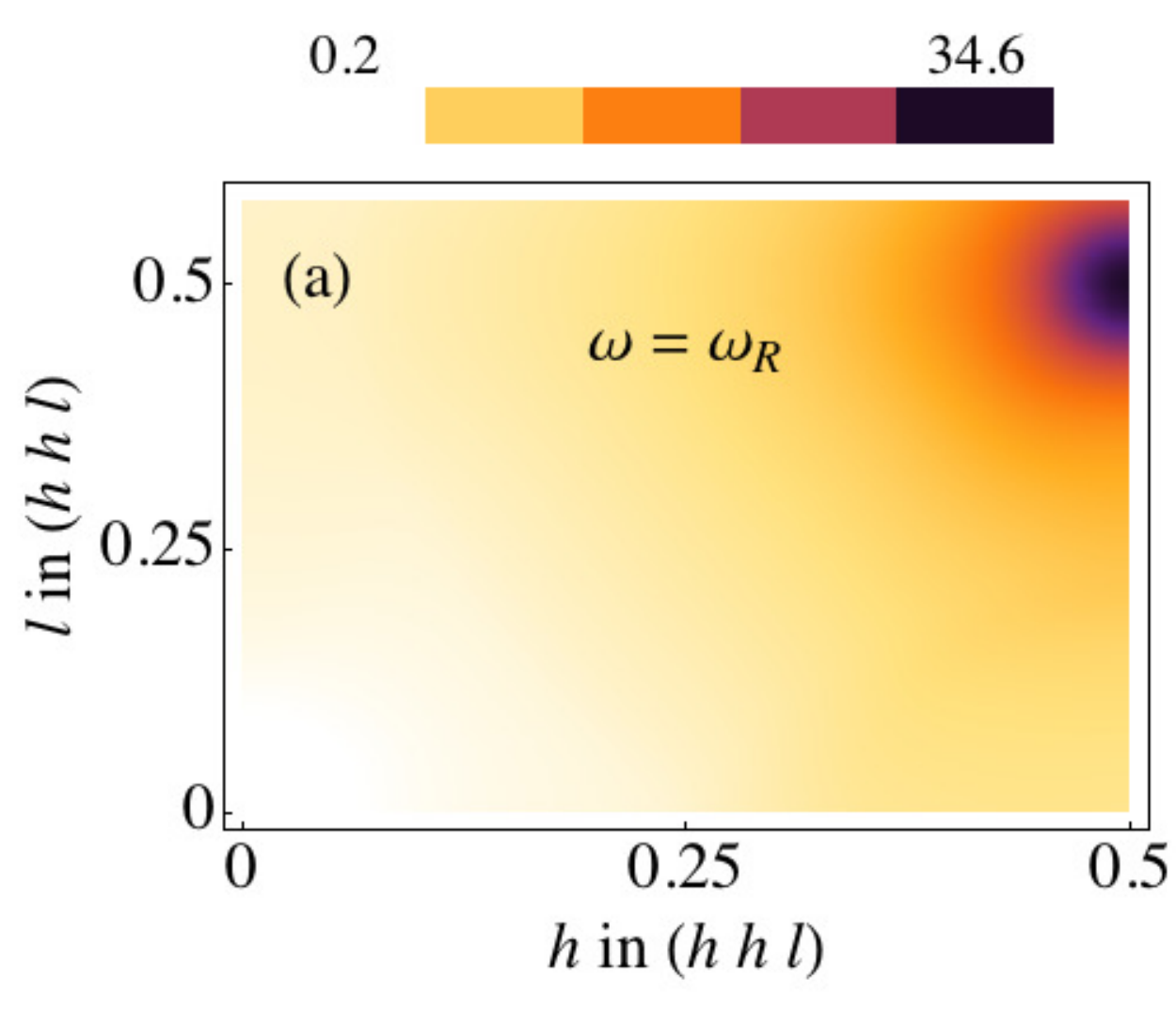}
\includegraphics[width=0.4\linewidth]{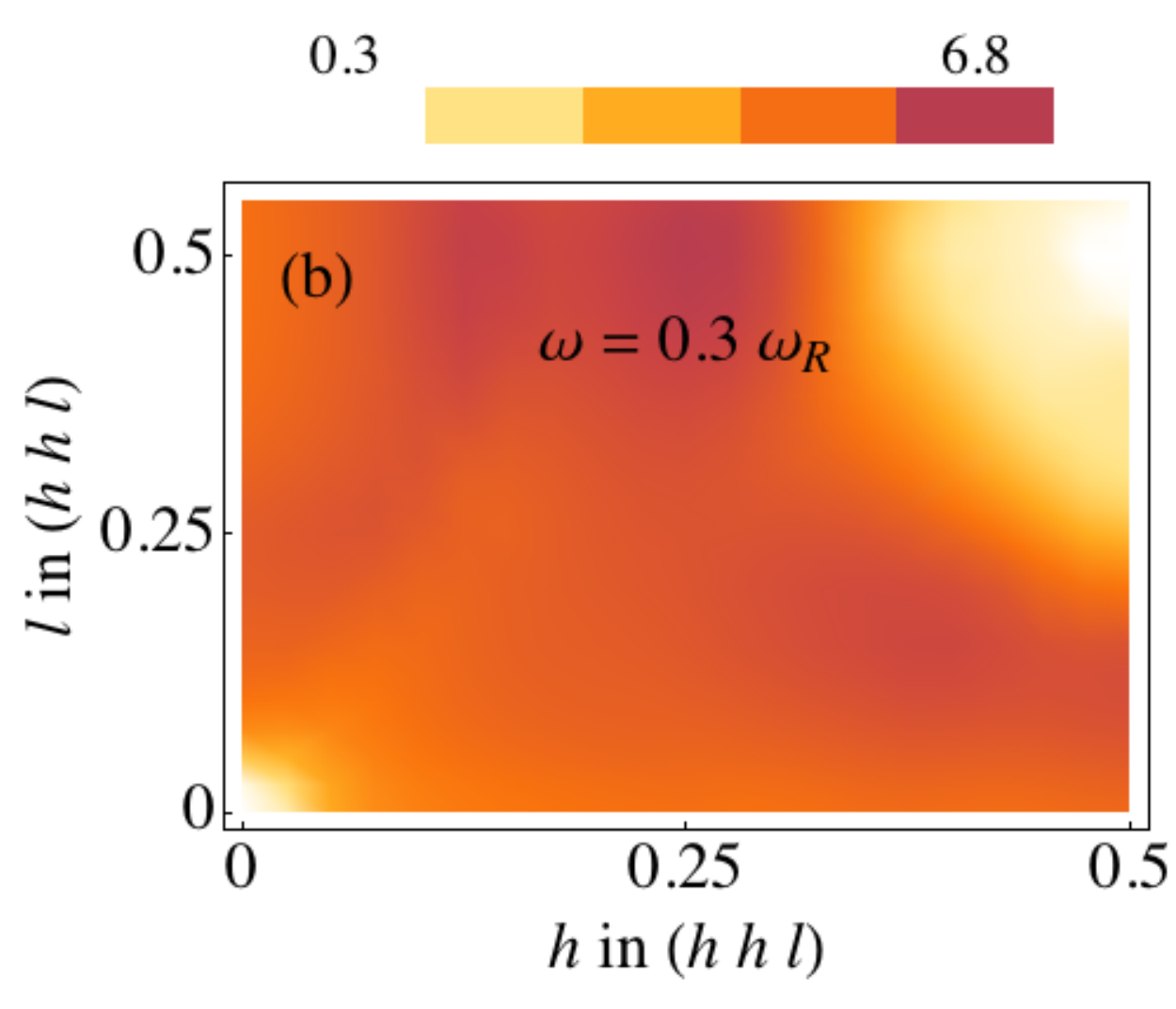}
}
\caption{
(Color online) Contour plot of imaginary part of RPA dynamical
susceptibility in (hhl)-plane of the reciprocal space; (a) at $\omega=\omega_r=0.07t$ (spin exciton resonance energy) pronounced
localized peak at R-point appears
(b) for energy in the spin gap, i.e.,  $\omega=0.3\omega_r$. Intensity at R-point vanishes due to spin gap formation.}
\label{fig:Fig3}
\end{figure}
%
A complementary constant-$\omega$ plot of the magnetic scattering intensity which is proportional to $\chi^"_{RPA}({\bf q},\omega=const)$ is shown in  Fig.~\ref{fig:Fig3} for \bq~ in the (hhl) plane as in the experimental scattering geometry.  At the resonance position $\omega_r$ (a) the momentum dependent scattering intensity is strongly peaked at the R-point with rapid decay in all \bq~ directions into the scattering plane. On the other hand for $\omega=0.3\omega_r$ (b) 
which is in the spin gap region, the latter shows up as a complete depletion of intensity at the R-point. Due to the magnetic sum rule the formation of the spin gap at this energy leads to a roughly even redistribution of the spectral weight across the whole scattering plane. This complete change of constant-$\omega$ intensity in (hhl) plane for $\omega=\omega_r$ and  $\omega\ll\omega_r$ is in agreement with the experimental result\cite{friemel:11}.

Now we discuss the temperature dependence of resonance  intensity. We start from itinerant type AFQ/AFM order parameters in Eq.~(\ref{eq:OP}) with a typical MF BCS temperature dependence shown in  Fig.~\ref{fig:Fig1}b (inset). The resonance intensity at the HO wave vector in Fig.~\ref{fig:Fig1}c appears already at T$_Q$ and is further enhanced below T$_N$. Experimentally it is found that it is strongly suppressed in the region $T_N<T<T_Q$. This is an effect of quadrupole OP fluctuations at zero field due to the near degeneracy with octupole order\cite{shiina:01} which strongly suppress its amplitude. For example specific heat jump $\Delta C(T_Q)$ for H=0 is almost absent\cite{hiroi:97} while $\Delta C(T_N)$ is pronounced. However in finite fields of a few Tesla the AFQ HO is stabilized and $\Delta C(T_Q,H)$ is strongly enhanced. The stabilization of $\Delta_{\bQ'}$ in field is also directly know from RXS experiments\cite{matsumura:09}. This effect will also be present for the dynamical resonance. We therefore predict that the resonance peak at R will appear already in the temperature range $T_N<T<T_Q$ when comparable fields are applied.  We note that  even in the case of a single superconducting order parameter  the temperature dependence of the intensity generally deviates from the BCS MF behaviour.

In summary the recent INS experiments\cite{friemel:11} require a  rethinking of the HO phenomena in \Ce. The appearance of an itinerant spin exciton resonance at the AFQ wave vector $\bQ' $ proves that the previous restriction to localized 4f states in \Ce~for the hidden AFQ order is oversimplified. The neglect of itinerant aspects can no longer be upheld. The theory presented here is therefore built on the delocalized heavy quasiparticle states. They are gapped due to the effect of hybridization and AFQ/AFM type particle-hole condensation leading to an enhanced magnetic response at the R-point. Due to quasiparticle interaction a pronounced spin exciton resonance at this wave vector appears. Its salient features of momentum, energy and temperature dependence are in agreement with experimental observation. Therefore \Ce~ is the first non-superconducting heavy fermion example with a spin exciton resonance excitation originating in the AFQ hiden order state.

We thank D. S. Inosov for communicating experimental results and  M. Siahatgar for useful discussions.

\bibliographystyle{prl}

\bibliography{CeB6}

\end{document}